\documentclass[%
 reprint,
 amsmath,amssymb,
 aps,
]{revtex4-1}
\newcommand{\cev}[1]{\reflectbox{\ensuremath{\vec{\reflectbox{\ensuremath{#1}}}}}}
\usepackage{graphicx}
\usepackage{xcolor}
\usepackage{dcolumn}
\usepackage{bm}

\begin{document}

\preprint{APS/123-QED}

\title{Geometrical properties of the ground state manifold in the spin boson model}
\author{Lo\"{i}c Henriet}
\address{ICFO-Institut de Ciencies Fotoniques, The Barcelona Institute of Science and Technology, 08860 Castelldefels (Barcelona), Spain}

\date{\today}

\begin{abstract}
Geometrical and topological properties of quantum ground state manifolds permits to characterize phases of matter, and identify phase transitions. Here, we study the effect of a quantum dissipative environment on the geometrical properties of the ground state manifold of a single spin 1/2 in an external effective magnetic field. We show that the quantum phase transition at zero temperature in the model is associated with a universal metric singularity related to the divergence of the spin susceptibility. The absence of transition at finite temperature corresponds to a smooth variation of the associated metric with temperature, without singular points.
\end{abstract}

\maketitle


\section{\label{sec:introduction} Introduction}

The study of topological phases of matter has motivated the characterization of the geometry and topology of quantum ground state manifolds \cite{Kolodrubetz13,Ma13}. These exotic phases are indeed characterized by a global topological index, the Chern number. This topological index can be expressed as the integral of the Berry curvature, a local geometrical quantity, over a closed manifold in momentum space. The Berry curvature corresponds in fact to the antisymmetric part of a more general quantum (or Fubini-Study) metric tensor, describing the local geometry of the quantum ground state manifold under the change of an external parameter. Past studies have investigated the relation between the divergence of the quantum metric tensor and quantum phase transitions \cite{zanardi07,Venuti07,Gu10,You07,Carollo05,Zhu06,Ma09}. Here, we study the quantum metric tensor in the spin boson model, describing the interaction of a single spin 1/2 with a bosonic environment of harmonic oscillators. In a certain range of parameters, this model is known to exhibit a quantum phase transition at zero temperature from a delocalized phase to a localized phase for the spin when increasing the coupling with the environment. We show that a component of the quantum metric displays a divergence at the quantum phase transition, which can be related to the divergence of the spin susceptibility. Studying the effect of temperature on the ground state manifold geometry requires the introduction of a metric in the space of density matrices, the Bures metric. This metric is intimately related to the construction of Uhlmann holonomy \cite{Uhlmann89}, generalizing the notion of geometrical phases from wave functions to density matrices. Experimental measures of Uhlmann geometric phase were reported in Refs. \cite{Zhu11,Viyuela18}, and recent works studied the behavior of related quantities in relation with finite temperature transitions \cite{Viyuela14,Huang14,Budich15,Mera17}. Here, we show that no singularity occurs for the Bures metric at non-zero temperature in the case of the ohmic spin-boson model.\\

The paper is structured as follows. In Sec. \ref{sec:model_and_properties}, we introduce the spin boson model and recall the main properties of the quantum phase transition in the model. We then turn in Sec. \ref{sec:qpt} to the study of the quantum metric divergence at the quantum phase transition. Sec. \ref{sec:increasing_T} is devoted to the investigation of finite temperature effects on the ground state manifold geometry. In particular, we demonstrate there that the Bures metric remains regular for all non-zero temperatures at the exactly solvable Toulouse point.\\

\section{\label{sec:model_and_properties} Spin boson model}

We consider in the following a spin-1/2 in a radial (effective) magnetic field with a dissipative coupling along the z-axis, corresponding to the Hamiltonian $\mathcal{H}=\mathcal{H}_{TLS}+\mathcal{H}_{diss}$, with (we take $\hbar=1$)
\begin{align}
\mathcal{H}_{TLS}&=-\frac{1}{2}\vec{d}.\vec{\sigma}, \label{Hisol}\\
\mathcal{H}_{diss}&= \sigma^z \sum_k \frac{\lambda_k}{2} (b_k +b_k^{\dagger})+ \sum_k \omega_k \left(b_k^{\dagger}b_k+\frac{1}{2}\right).
\label{Hdiss}
\end{align}
where $\vec{\sigma}=(\sigma^x,\sigma^y,\sigma^z)$ and $\vec{d}=d( \sin \theta \cos \phi ,  \sin \theta \sin \phi, \cos \theta )^T$ is an external effective magnetic field. Here, $d>0$ is the norm of the magnetic field, $\theta$ and $\phi$ are respectively the polar and azimuthal angle of the magnetic field. $\mathcal{H}_{diss}$ is a microscopic model of the dissipation acting on the system \cite{weiss,leggett1981,leggett:RMP}. $b_k$ ($b_k^{\dagger}$) corresponds to the annihilation (creation) operator of a bosonic mode $k$ with frequency $\omega_k$. The interaction between the spin and the environment is fully characterized by the spectral function $J(\omega)=\pi \sum_k \lambda_k^2 \delta(\omega-\omega_k)$. In the following, we assume that $J$ is a smooth function of the frequency, $J(\omega)=2 \pi \alpha \omega^s \omega_c^{1-s} \exp \left(-\omega/\omega_c\right)$ with $s>0$. $\omega_c \gg d$ denotes a high frequency cutoff in the model, and $\alpha$ quantifies the strength of the spin-bath interaction.\\

In the remainder of the article, we will mainly focus on the case of an ohmic spectral function characterized by $s=1$. In this case, the coupling with the bath induces (for $\theta=\pi/2$) a quantum phase transition from a delocalized phase for the spin, characterized by $\langle \sigma^z \rangle=0$, to a localized phase with $\langle \sigma^z \rangle=\pm 1$ (see for example Refs. \cite{leggett:RMP,weiss,Hur}). In the scaling regime defined by $d/\omega_c \ll 1$, the critical value of the coupling is $\alpha_c\simeq 1$, and this transition belongs to the Kosterlitz-Thouless universality class. We will also briefly generalize our results to the case of sub-ohmic spectral densities, with $0<s<1$, where a second order phase quantum phase transition occurs with a critical coupling $\alpha_c<1$.


\section{\label{sec:qpt} Quantum phase transition and quantum metric singularity at zero temperature}

Let us consider a family of quantum states $| \psi (\lambda)\rangle$ in an Hilbert space $H$, parametrized by a continuous external parameter $\lambda$. A natural measure of distance between $| \psi (\lambda_1)\rangle$ and $| \psi(\lambda_2)\rangle$ is given by $\mathcal{N}(|\psi(\lambda_1)\rangle-|\psi(\lambda_2)\rangle)$, where $\mathcal{N}$ denotes the usual Hilbert space norm. The distance between two wave functions corresponding to a infinitesimal change in parameter space $d\lambda$ is then given by 
\begin{align}
ds^2= g_{\mu \nu} d\lambda^{\mu} d\lambda^{\nu}\label{line_element_fubini},
\end{align}
where $g_{\mu \nu}=\textrm{Re} \left[\chi_{\mu \nu} \right]$ are the elements of the Fubini-Study metric \cite{provost1980}, with
\begin{align}
\chi_{\mu \nu}=\langle \psi|\cev{\partial}_{\lambda^{\mu}}\partial_{\lambda^{\nu}}|\psi\rangle-\langle \psi|\cev{\partial}_{\lambda^{\mu}}|\psi\rangle\langle \psi|\partial_{\lambda^{\nu}}|\psi\rangle \label{metric_elements}.
\end{align}
In Eqs. (\ref{line_element_fubini}) and (\ref{metric_elements}), the summation over indices is implicit and $\langle \psi|\cev{\partial}=\partial\langle \psi|$. The antisymmetric part of the quantum metric tensor $\chi_{\mu \nu}$ defines the Berry curvature $\mathcal{F}_{\mu \nu}$. We are here interested in the case where the state $|\psi \rangle$ above corresponds to the ground state wave function $|g\rangle$ of the spin boson model at zero temperature, and the external parameters are the radial and azimuthal angles, i.e. $\lambda=(\theta,\phi)$. For an isolated spin 1/2, i.e. $\alpha=0$, one easily finds the components of the metric. In particular, we get $g_{\theta \theta}=1/4$.\\

At non-zero coupling with the environment, one can generally write the ground state of the Hamiltonian $\mathcal{H}$ under the form 
\begin{align}
|g\rangle =\frac{1}{\sqrt{p^2+q^2}} \left[p(\theta) e^{-i\phi} |\uparrow\rangle \otimes |\chi_{\uparrow}\rangle+q(\theta)|\downarrow\rangle \otimes |\chi_{\downarrow}\rangle\right],\label{spin_state}
\end{align}
where $p$ and $q$ are two real numbers, and $|\chi_{\sigma}\rangle$ are bath states associated with the spin polarization $\sigma=\uparrow,\downarrow$ direction. Determining $|\chi_{\sigma}\rangle$, as well as the coefficients $p$ and $q$, for an arbitrary value of the angles $\theta$ and $\phi$ is in general difficult. \\

When the coupling with the external environment is weak ($\alpha \ll 1$), one can treat the bosonic bath in a perturbative manner to determine $|g\rangle$, and in turn the metric elements of Eq. (\ref{metric_elements}). One example corresponds to the single- or multi-polaron variational approximation \cite{bera:PRB} (see also Refs. \onlinecite{Silbey_Harris,leggett:RMP,weiss,Hur}), that we use in the next subsection. This will allow us to characterize the deformation induced by the bosonic bath on the geometry of the ground state manifold. We will see in particular that it induces a $\theta$-dependency for $g_{\theta \theta}$, which becomes more peaked around $\theta=\pi/2$. The polaron picture will allow us to understand physically the origin of this change.

\subsection{Shifted oscillators}
  
  \begin{figure}[h!]
\center
\includegraphics[scale=0.42]{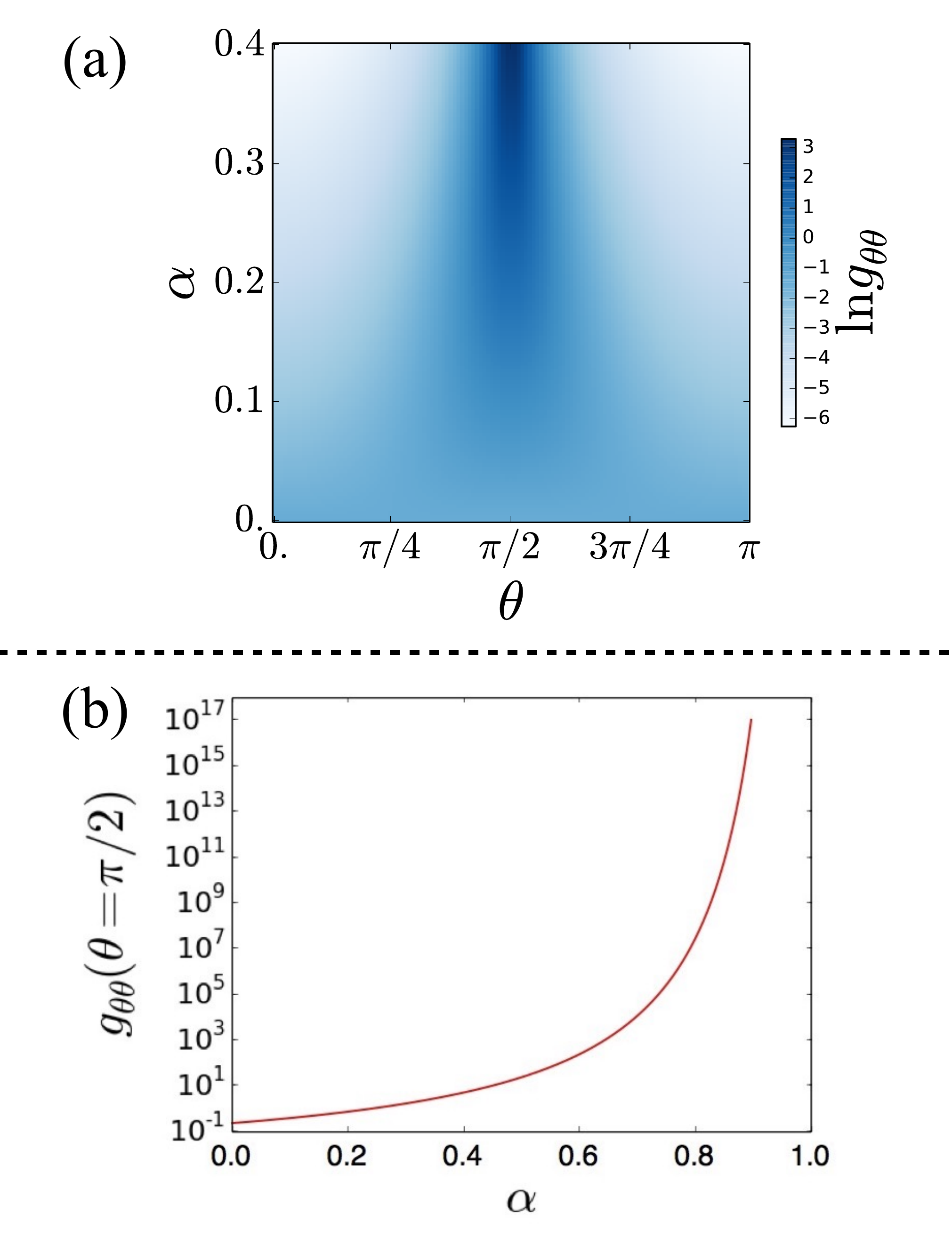}
\caption{\label{Metric_polaron} (a) Evolution of the metric element $g_{\theta \theta}$ with respect to the angle $\theta$ and the dissipation parameter $\alpha$, evaluated with the single-polaron ansatz. (b) Evolution of the metric component $g_{\theta \theta}$ at the equator $\theta=\pi/2$ with the dissipative parameter $\alpha$. These results correspond to a ohmic bath $s=1$ with $\omega_c/d=10$. }
\end{figure}

 At weak dissipation, one can approximate bath states $|\chi_{\sigma}\rangle$ by multi-mode coherent states of the form $|\chi_{\sigma}\rangle=\exp\left[\sum_k f_k^{\sigma} (b_k -b_k^{\dagger}) \right]|0\rangle$, where $|0\rangle$ denotes the vacuum for all oscillators. In this polaron picture, each bath mode is supposed to be in the ground state of an harmonic oscillator whose equilibrium position has been shifted by the amount $f_k^{\sigma}$. The set of real numbers $\{f_k^{\uparrow },f_k^{\downarrow },p\}$ are then found variationally by minimizing the mean value of the energy, $E=\langle g | \mathcal{H} |g\rangle$ of the system. This gives access to the values of $p(\theta)$ and $q(\theta)$ at weak coupling, allowing us to determine the metric elements. The approximate results of this approach are shown in Fig. \ref{Metric_polaron} (a), where we plot the evolution of $g_{\theta \theta}$ with respect to $\theta$ and the dissipation stength $\alpha$ in the case of an ohmic bath $s=1$ with high frequency cutoff $\omega_c/d=10$. At $\alpha=0$, we check that the metric element is uniform with respect to $\theta$. At non-zero coupling strength $\alpha$, ones notices a progressive increase of the metric element around $\theta=\pi/2$. This effect can be understood by evaluating the retroaction of bath states on the spin. One can indeed note that bath states tend to re-inforce the polarization of the spin along the z-axis. More specifically, the non-zero coherence of shifted bath states induces an additional effective field $-h_{ind} \sigma_z$ along the z-axis for the spin, with $h_{ind}>0$ ($h_{ind}>0$) for $\theta<\pi/2$ ($\theta>\pi/2$). As a result of this retroaction, one has $\langle \sigma^z \rangle>\langle \sigma^z \rangle_0$ ($\langle \sigma^z \rangle<\langle \sigma^z \rangle_0$) for $\theta<\pi/2$ ($\theta>\pi/2$), where $\langle \sigma^z \rangle_0$ refers to the spin-expectation value of the isolated system without any coupling to the environment ($\alpha=0$). This in turn leads to a strong variation of $p$ and $q$ with respect to $\theta$ close to $\theta=\pi/2$ where $\langle \sigma^z \rangle$ changes sign and a corresponding increase of $g_{\theta \theta}$.





\begin{figure}[h!]
\center
\includegraphics[scale=0.6]{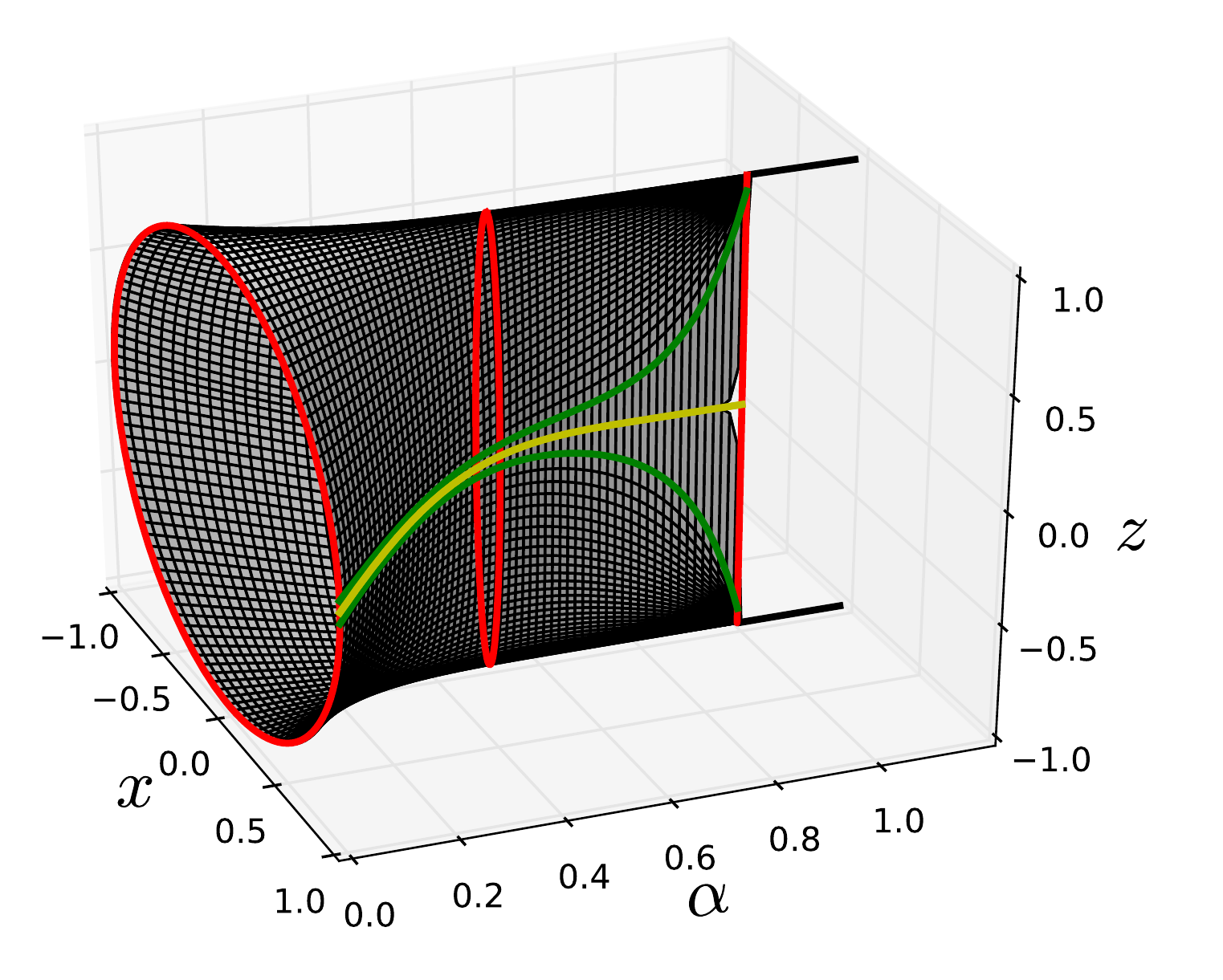}
\caption{\label{Deformation_metric_alpha} Sketch of the deformation of a section of the Bloch sphere with dissipation strength $\alpha$ for the ohmic spin boson model. The red lines show the section $(x,z)$ for $\alpha=0$, $\alpha=0.5$ and $\alpha=\alpha_c^- \sim 1$, which go from a circle at $\alpha=0$ to a line at $\alpha=\alpha_c^-$.  The yellow line corresponds to the evolution of $(x,z)$ with $\alpha$ for a fixed value $\theta=\pi/2$. The green lines show the evolution of $(x,z)$ with $\alpha$ for two particular values, $\theta=\pi/2 \pm u$ with a small $u>0$. For these two values of $\theta\neq \pi/2$, $\langle \sigma^z \rangle$ and $\partial_{\alpha}\langle \sigma^z \rangle$ are continuous with $\alpha$ \cite{Hur}.}
\end{figure}

\subsection{Exact result $\theta=\pi/2$}
The increase of the curvature close to $\theta=\pi/2$ identified previously can be more precisely described. From Eqs. (\ref{metric_elements}) and (\ref{spin_state}), one more specifically finds that the metric element $g_{\theta \theta}$ is related to the spin susceptibility at $\theta=\pi/2$,
 \begin{align}
g_{\theta \theta} (\theta=\pi/2) =\left[ \frac{\partial_{\theta}\langle \sigma^z \rangle (\theta=\pi/2)}{2} \right]^2.\label{metric_equator}
\end{align}
One therefore expects a divergence of the metric for $\theta=\pi/2$ when $\alpha$ approaches the critical coupling $\alpha_c$ in the cases of ohmic and sub ohmic baths. This divergence follows from the behavior of the spin susceptibility at quantum criticality. In the case of an ohmic bath, exact Bethe-Ansatz results \cite{Bethe_1,Bethe_2,Hur} allow us to determine the evolution of $\langle \sigma^z \rangle$ with respect to $\theta$, leading to $g_{\theta \theta} (\theta=\pi/2) \propto \left(\frac{\omega_c}{d}\right)^{\frac{2\alpha}{1-\alpha}}$ for $\alpha<1$. The evolution of this diverging metric element with $\alpha$ is shown in Fig. \ref{Metric_polaron} (b) in this precise case of ohmic dissipation.  In the case of a sub-ohmic bath, the behavior of the spin susceptibility at criticality \cite{Vojta05} gives $g_{\theta \theta} (\theta=\pi/2) \propto (\alpha_c-\alpha)^{-2\gamma}$, with $\gamma=1+\mathcal{O}(s)$. These results generalize the previous study of Ref. \cite{henrietthese2016,henriet17,toporeview}, where it was shown that the Berry curvature diverges at the quantum phase transition, with $\mathcal{F}_{\phi \theta} \propto  \partial_{\theta}\langle \sigma^z \rangle (\theta=\pi/2)/2$. The deformation of the ground state manifold with increasing dissipation, and the corresponding metric singularity at the critical coupling, can be conveniently visualized by examining the surface described by the equilibrium Bloch vector $(x,y,z)=(\langle \sigma^x \rangle, \langle \sigma^y \rangle, \langle \sigma^z \rangle)$ when varying $\theta$ and $\phi$. For a coupling along the z-axis as in Eq. (\ref{Hdiss}), such a surface is invariant under $\phi$-rotations and it is therefore equivalent to study sections of constant $\phi$. We show in Fig. \ref{Deformation_metric_alpha} a sketch of the evolution of the $\phi=0$ section with the dissipation parameter $\alpha$, in the case of ohmic dissipation. At $\alpha=0$, this section corresponds to a circle. This circle is progressively compressed along the x-direction as $\alpha$ increases. When $\alpha>1$, the only possible equilibrium Bloch vector correspond to $\langle \sigma^z \rangle=\pm 1$, i.e. the north and south poles of the Bloch sphere. Crossing the quantum phase transition point therefore corresponds to a breaking of the surface, with no continuous path between the possible Bloch vectors for a given value of $\alpha>\alpha_c$. It is important to note that the quantum phase transition in the model is restricted to the point $\theta=\pi/2~[\pi]$. For any non-zero value of the field along the $z$-direction, increasing the coupling $\alpha$ leads to a crossover \cite{Hur}. This can be visualized in Fig. \ref{Deformation_metric_alpha}, by following the green lines illustrating the evolution with $\alpha$ of the curve $(x,z)(\theta_1)$ for an angle $\theta_1=\pi/2 \pm u$, with a small value of $u>0$.

\section{\label{sec:increasing_T} Absence of metric singularity at nonzero temperature}

We have shown in the previous section that increasing the coupling strength $\alpha$ induces a divergence of the Fubini-Study metric at the quantum phase transition. In this Section, we now study the effect of the temperature on the geometry on the ground state manifold. At non-zero temperature, one can no longer use the Fubini-Study metric of Eq. (\ref{metric_elements}) as the system cannot be described by a pure state. One needs instead to work with the density matrix of the system, which is an operator acting on the Hilbert space $H$. The matrix algebra of operators acting on $H$ is a vector space called the Hilbert-Schmidt vector space $HS$. The natural metric on the space of density matrices is the Bures metric, which is obtained from the construction of the Hilbert-Schmidt bundle from state purification \cite{bengtsson}. For the density matrix of a single spin 1/2, explicit results can be obtained and one finds that the Bures line element squared of the density matrix $\rho$ 
\begin{align}
\rho =\frac{1}{2}\begin{bmatrix}
    1+z       & x-iy \\
    x+iy       & 1-z
\end{bmatrix}\label{density_matrix}
\end{align}
is given by
\begin{align}
ds^2=\frac{1}{4}\left[dx^2+dy^2+dz^2+\frac{(x dx+y dy+zdz)^2}{1-x^2-y^2-z^2} \right].\label{Bures_distance}
\end{align}


One then recovers that the element $g^B_{\theta\theta}$ of the Bures metric, at $\theta=\pi/2$ where $z=\langle \sigma^z \rangle=0$ and $\partial_{\theta} x=\partial_{\theta} \langle \sigma^x \rangle=0$ to be $g^B_{\theta\theta}(\theta=\pi/2) =1/4\left[ \partial_{\theta}\langle \sigma^z \rangle (\theta=\pi/2) \right]^2$. At non-zero temperature, the divergence of the spin susceptibility is washed out and no quantum phase transition occurs. As a result, no metric singularity occurs for the ohmic spin boson model when increasing the temperature for a fixed value of the coupling $\alpha<\alpha_c$. In the next Section, we study more precisely the evolution of the metric with temperature at the exactly solvable Toulouse point $\alpha_T=1/2$.

\subsection{\label{sec:Toulouse} Exact results at the Toulouse point, $\alpha_T=1/2$}

\begin{figure}[h!]
\center
\includegraphics[scale=0.4]{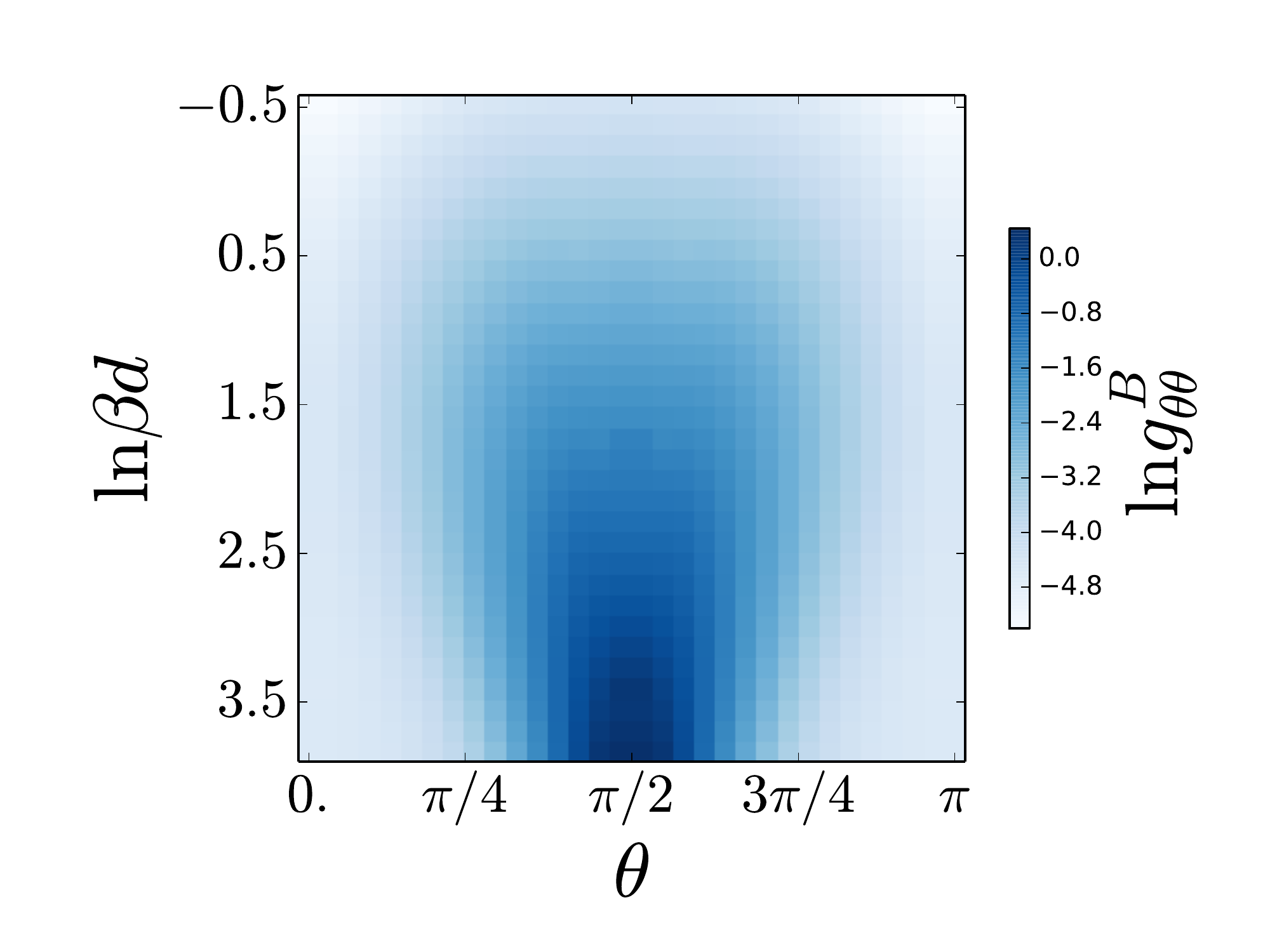}
\caption{\label{Bures_toulouse} Evolution of the Bures metric $g^B_{\theta\theta}$ as a function of $\theta$ and $\beta d$, for the ohmic spin boson model at the exactly solvable Toulouse point $\alpha_T=1/2$. We took $\omega_c/d=10$.}
\end{figure}

\begin{figure}[h!]
\center
\includegraphics[scale=0.13]{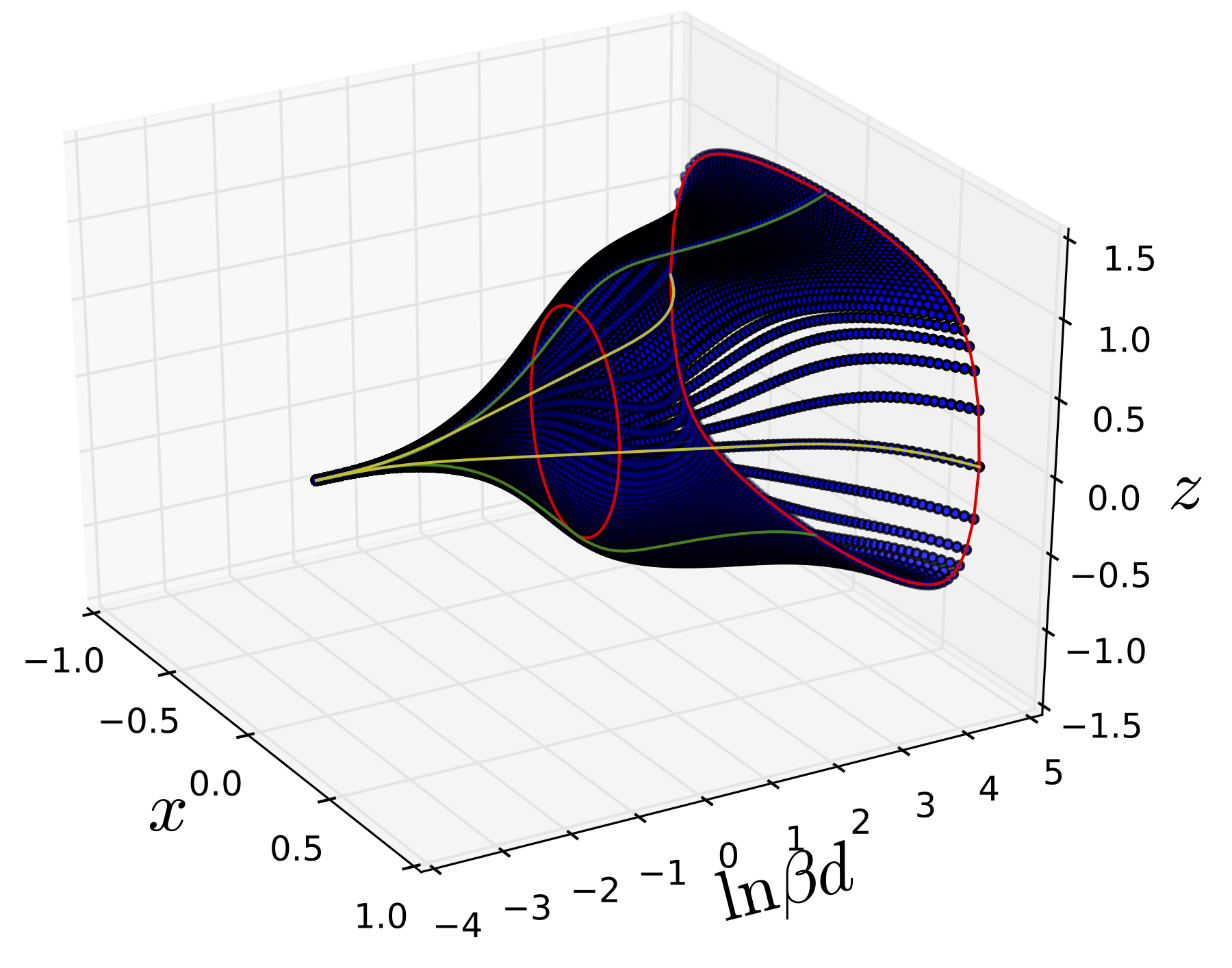}
\caption{\label{Bloch_section_toulouse} Deformation of the $\phi=0$ section of the surface described by the Bloch vector, as a function of $\beta d$, for the ohmic spin boson model at the exactly solvable Toulouse point $\alpha_T=1/2$. We took $\omega_c/d=10$. The two red lines correspond to two sections at particular values of the inverse temperature. The green and yellow lines correspond to particular values of $\theta=0 ~[\pi]$ and $\theta=\pi/2~ [\pi]$, respectively. }
\end{figure}

At the special point $\alpha_T=1/2$, the ohmic  spin-boson  Hamiltonian  can  be
mapped onto the exactly solvable non-interacting resonance level Hamiltonian \cite{Toulouse69,Anderson70,Guinea_bosonization,leggett:RMP,weiss}. In particular, the partition function is given at inverse temperature $\beta$ by \cite{weiss} 
\begin{align}
\mathcal{Z}=2\exp\left[\int_{x_0}^{\infty} dx \frac{1-e^{-4x\omega_c/(\beta d^2 \sin^2\theta)}\cos\left[d \cos \theta \beta  x/\pi\right]}{x \sinh x}\right]
\label{partition_function_Toulouse}
\end{align}
with $x_0=\pi/(\beta \omega_c)$. Equation (\ref{partition_function_Toulouse}) allows us to compute numerically the spin reduced density matrix at non zero-temperature and its evolution with the angle $\theta$. We show in Fig. \ref{Bures_toulouse} the evolution of the component of the Bures metric $g^B_{\theta\theta}$  with $\theta$ and $\beta d$. At low temperatures, we see that this metric component has a maximal value for $\theta=\pi/2$, in accordance with our analysis of the previous section. At higher temperatures, $g^B_{\theta\theta}(\theta=\pi/2)$ is reduced and the metric becomes more homogeneous, excluding as expected a possible divergence.  This behavior can notably visualized by studying how the $\phi=0$ section of the surface described by the Bloch vector evolves when increasing the temperature, which is shown in Fig. \ref{Bloch_section_toulouse}. At low temperatures, this section corresponds to a deformed circle. When the temperature increases, it is progressively transformed in a continuous manner and finally reaches a single point at $(\langle\sigma^x\rangle,\langle\sigma^z\rangle)=(0,0)$ at infinite temperatures. At infinite temperatures, all the metric elements tend to zero as the system is always at the Bloch sphere center for any value of $\theta$ or $\phi$. This behavior is in sharp contrast with the one illustrated in Fig. \ref{Deformation_metric_alpha}. Here, one witnesses a smooth and continuous modification of this section, in the absence of finite temperature transition. Note that the mapping mentioned here is valid in the regime $d/\omega_c \ll 1$.




\section{\label{sec:geodesic_def} Discussion and conclusions}
In this paper, we studied how the presence of a quantum dissipative environment affects the geometrical properties of the ground state manifold of a spin 1/2, both at zero and finite temperatures. We described these geometrical properties by using the Fubini-Study and the Bures metric. We showed that the Fubini-Study metric exhibits a singularity at the quantum phase transition in the case of ohmic and subohmic spectral functions, with a divergence characterized by the universality class of the transition. No singularity was identified for the Bures metric at non-zero temperatures. Our study confirms that metric tensors display universal behaviors in the vicinity of phase transitions.\\

This work has benefited from useful discussions with K. Le Hur and P. Orth, and conversations at the Eniqma meeting in Lille. The author acknowledges support from the Ministry of Economy and Competitiveness of Spain through
the “Severo Ochoa” program for Centres of Excellence in
Research and Development (SEV-2015- 0522), Fundaci\'{o} Privada Cellex, Fundaci\'{o}
Privada Mir-Puig, and Generalitat de Catalunya
through the CERCA program.


\bibliography{refs}
\end{document}